\documentstyle[aps,amsmath,multicol,prl,epsfig]{revtex}

\newcommand{\beq}{\begin{equation}}

\newcommand{\eeq}{\end{equation}}

\begin{document}

\begin{title}
{\bf On $3+1$ Dimensional Friedman-Robertson-Walker Universes with Matter}
\end{title}

\author{T. Christodoulakis$^1$\thanks{tchris@cc.uoa.gr}, C. Helias$^1$, P.G. Kevrekidis$^{2}$, I.G. Kevrekidis$^{3}$ and 
G.O. Papadopoulos$^1$\thanks{gpapado@cc.uoa.gr}}
\address{
$^{1}$ Nuclear and Particle Physics Section, Physics Department,
University of Athens,
Panepistimiopolis, Ilisia, Athens 15771, Greece \\
$^{2}$ Department of Mathematics and Statistics,University of Massachusetts,
Amherst MA 01003-4515, USA \\
$^3$ Department of Chemical Engineering, Princeton University,
6 Olden Str. Princeton, NJ 08544}

\maketitle

\begin{abstract}
We examine the dynamical behavior of matter coupled to gravity in
the context of a linear Klein-Gordon equation coupled to a
Friedman-Robertson-Walker metric. The resulting ordinary
differential equations can be decoupled, the effect of gravity
being traced in rendering the equation for the scalar field
nonlinear. We obtain regular (in the massless case) and asymptotic
(in the massive case) solutions for the resulting matter field and
discuss their ensuing finite time blowup in the light of earlier
findings. Finally, some potentially interesting connections of
these blowups with features of focusing in the theory of nonlinear
partial differential equations are outlined, suggesting the
potential relevance of a nonlinear theory of quantum cosmology.
\end{abstract}


\vspace{2mm}

In the past two decades there has been an extensive interest
in 3-dimensional gravity, particular after the demonstration of
the fact that its quantum version is solvable \cite{witten} and
that it contains black hole solutions \cite{banados}.

The idea of examining cosmological models in this context was, however,
of interest before \cite{gidd,burd,corn}, as well as after
\cite{chopt,cruz} these findings. In most of these studies
\cite{gidd,burd,corn,cruz}, the Friedman-Robertson-Walker (FRW)
metric was used and the resulting equations were ordinary
differential equation governing the time-evolution of the
scale factor of the relevant metric and the evolution of
the matter and/or radiation field coupled to it.

On a slightly different track, one can list the works of
\cite{chopt,koike,abraham} (see also references therein
and the review of \cite{gundl}), where the metric scale
factors were allowed to be temporally {\it as well as} spatially
variable, and the resulting partial-differential equations (PDEs)
were studied to obtain collapse type solutions.

In this brief report, we will restrict ourselves to the former
type of considerations (but in a 3+1 dimensional setting), namely
fixing the spatial dependence of the metric tensor to be of  the
FRW type (in co-moving coordinates and with a ``cosmological'' time
choice)
\begin{equation}
ds^2=-dt^2+a^2(t) \left( \frac{1}{1-k r^2} dr^2 + r^2 d\theta^2 +
r^2 \sin^2 \theta d \phi^2 \right)
\label{ceq1}
\end{equation}
and coupling gravity to the simplest possible model for matter,
namely a linear Klein-Gordon (KG) type equation for a massive
scalar. In the metric of Eq. (\ref{ceq1}), $a$ is the scale factor
while $k$ describes the curvature of the spatial slice and can be
normalized to the values $-1,0,1$ in the hyperbolic, flat and
elliptic case respectively. Previously such models were examined
in a more complicated setting (most often in 2+1 dimensions) where
either equations of state \cite{cruz} or special scalar field
potentials \cite{burd} were used to obtain closed form solutions.

Here, we will examine the simplest possible (physically relevant)
scalar field potential $V(\phi)=m^2 \phi^2/2$, which gives rise to
a linear (in $\phi$) equation for the scalar coupled to gravity.
In particular the Einstein-Klein-Gordon field equations in this
case read:
\begin{eqnarray}
-3 \frac{k+\dot{a}^2}{a^2}+ \rho_{\phi}=0
\label{ceq2}
\\
\frac{k+\dot{a}^2+2 a \ddot{a}}{a^2}+ p_{\phi}=0
\label{ceq3}
\\
\ddot{\phi}+m^2 \phi+\frac{3 \dot{a}}{a} \phi=0
\label{ceq4}
\end{eqnarray}
where the energy density and the pressure associated with
the scalar field are given respectively by (see e.g., \cite{cruz})
\begin{eqnarray}
\rho_{\phi}=\frac{1}{2} ( \dot{\phi}^2 + m^2 \phi^2) \equiv \frac{1}{2}
\dot{\phi}^2 + V(\phi),
\label{ceq5}
\\
p_{\phi}=\frac{1}{2} ( \dot{\phi}^2 - m^2 \phi^2) \equiv \frac{1}{2}
\dot{\phi}^2 - V(\phi).
\label{ceq6}
\end{eqnarray}
Note that the first is the quadratic constraint
$G^{0}_{0}=T^{0}_{0}$; the second is the only independent spatial
equation $G^{1}_{1}=T^{1}_{1}$ while the last is the dynamical
equation for the scalar field, i.e., the Klein-Gordon
equation. $G$ and $T$ denote the Einstein curvature tensor and
the energy-momentum tensor respectively.

In the above equations the dot denotes temporal derivative.
Naturally, Eqs. (\ref{ceq2})-(\ref{ceq4}) are not linearly
independent as the linear combination of the derivative of the
first and of the second can be used to obtain the third.

We first examine the spatially flat case of $k=0$, and use it as a
guide. In this setting one immediately observes that from Eqs.
(\ref{ceq2}) and (\ref{ceq4}), two separate expressions for
$\dot{a}/a$ can be obtained, hence equating them, a second-order
ordinary differential equation (ODE) emerges for the scalar field
$\phi$ in the form:
\begin{eqnarray}
\frac{3}{2} \dot{\phi}^2 \left(m^2 \phi^2 + \dot{\phi}^2 \right)=
\left(m^2 \phi + \ddot{\phi} \right)^2.
\label{ceq7}
\end{eqnarray}
 Notice that this is the only case (among the ones that we will
examine) in which the resulting ODE is of 2nd order. In the
remaining cases, the ODE is of 3rd order. Moreover, it is worth
commenting on the nature of this ODE: in particular, the resulting
equation is {\it nonlinear}. Hence, even though the inclusion of
matter is realized through a linear dynamical equation, in the
setting of even the simplest cosmological models, the coupling of
matter to gravity induces the emergence of a nonlinear equation;
it is as if the ``trace'' of gravity, when the equation for the
scalar field is decoupled from it, remains in the nonlinearity of
the resulting ODE.

As the simplest possible case among the ones with $k=0$,
we examine the massless case, i.e., $m=0$. In the latter
setting, we obtain the solutions (up to a constant shift)
of the form:
\begin{eqnarray}
\phi=\pm \sqrt{\frac{2}{3}}  \log\left( t^{\star}-t \right)
\label{ceq8}
\end{eqnarray}
which in turn results in a power law dependence of the scale
factor of the form $a \sim (t^{\star}-t)^{1/3}$. Hence, a blowup
(focusing) type effect occurs at $t=t^{\star}$ in a logarithmic
fashion for the scalar and the corresponding scale factor shrinks
to 0 with a power law dependence. Such dependencies are 
reminiscent of the critical blowup in prototypical nonlinear PDEs
with focusing solutions (for a spatio-temporally dependent field
$\psi$) such as the nonlinear Schr{\"o}dinger equation
\cite{landman,sulem,skk}
\begin{eqnarray}
i \psi_t=- \Delta \psi - |\psi|^{2 \sigma} \psi.
\label{ceq9}
\end{eqnarray}
$\Delta$ stands for the Laplacian and $\sigma$ the power of the
nonlinearity. For $d \sigma<2$, no focusing solutions occur (d is
the dimensionality of the Laplacian); when $d \sigma=2$,
logarithmically slow focusing phenomena take place (in the
corresponding ``proper time'' see \cite{landman,sulem,skk}), while
when $d \sigma >2$, the so-called strong collapse occurs, where
the focusing happens with a power law dependence \cite{sulem,skk}.
Hence, the {\it critical} case of the nonlinear focusing phenomena
in equation (\ref{ceq9}) shares some of the collapse
characteristics of the present model.

\begin{figure}[tbp]
\epsfxsize=7.5cm
\centerline{\epsffile{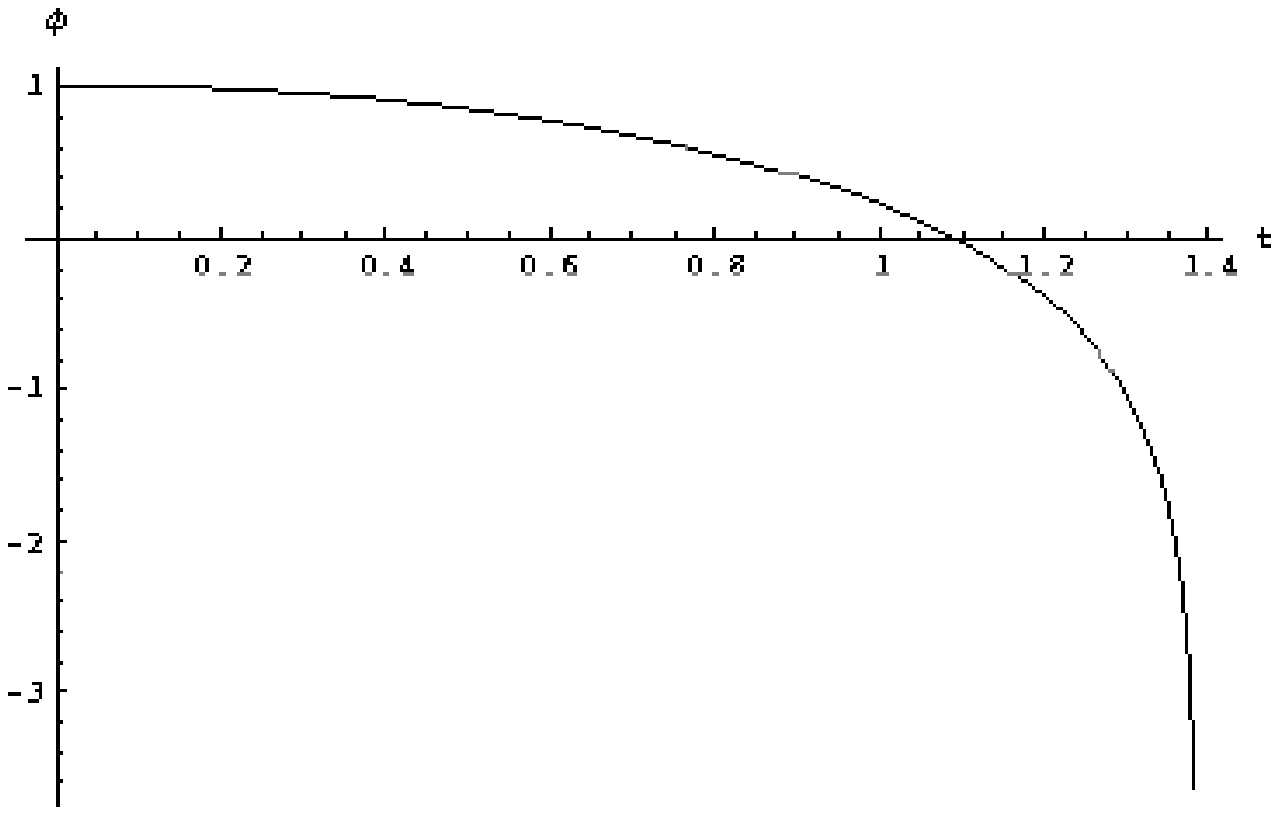}}
\caption{Logarithmic blowup of the scalar field as a function
of time in the massive case (for $m=1$), starting with $\phi(0)=1$,
$\dot{\phi}(0)=0$. The blowup occurs for $t \approx 1.38$. }
\label{fmfig1}
\end{figure}

We now turn to the massive case (still for $k=0$). In the latter,
we can no longer solve the problem analytically. However, we can
directly observe that if we adopt a solution of logarithmic
dependence of the form of Eq. (\ref{ceq8}) in this case as well,
the massive terms diverge much more slowly (at worst as
$(\log(t^{\star}-t))^2/(t^{\star}-t)^2$, as opposed to the
$(t^{\star}-t)^{-4}$ divergence of the dominant (massless) terms).
This signifies that asymptotic self-similarity will ensue from
this case and the logarithmic dependence will eventually set in
and dominate the asymptotic behavior leading to collapse. An
example of this is shown in Fig. \ref{fmfig1}, where a numerical
simulation of Eq. (\ref{ceq7}) clearly indicates the logarithmic
focusing of the scalar field for as $t \rightarrow 1.38$. This
asymptotic type of self-similarity occurs quite often in
cosmology, as can be seen for example in \cite{review} (and
references therein). One final comment worth making about this
case is that it appears as if the massive case is asymptoting to
the results for the massless one.

\begin{figure}[tbp]
\epsfxsize=7.5cm
\centerline{\epsffile{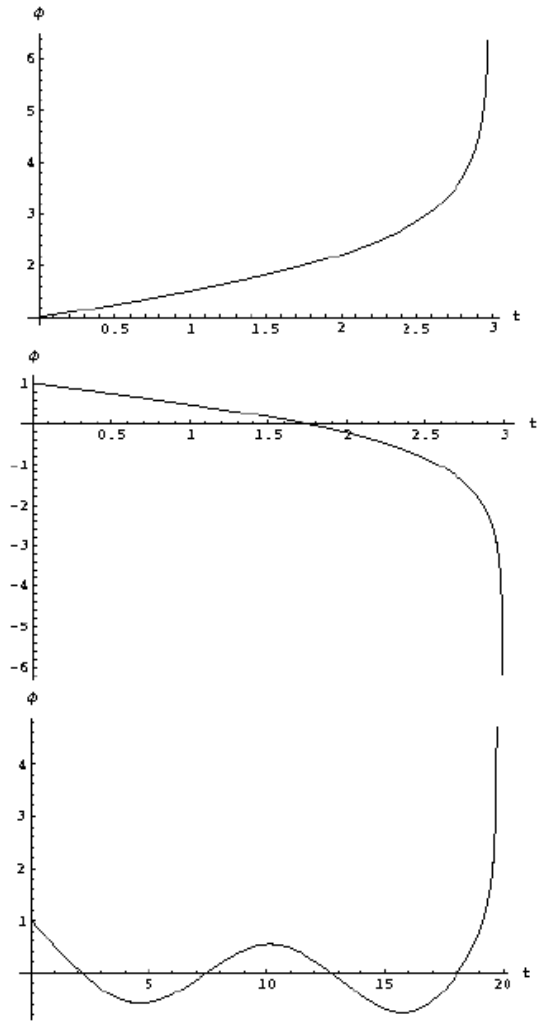}}
\caption{Logarithmic blowup of the scalar field as a function
of time in the massive case (for $m=0.1$), starting with $\phi(0)=1$,
$\dot{\phi}(0)=0.5$, $\ddot{\phi}(0)=0$ (top panel) and
$\phi(0)=1$, $\dot{\phi}(0)=-0.5$, $\ddot{\phi}(0)=0$ (middle panel).
The blowup occurs for $t \approx 3$ in these cases. A case with
a larger mass ($m=0.6$) is shown in the bottom panel. In this
case, the blowup occurs at later times (in this case for $t \approx 20$).}
\label{fmfig2}
\end{figure}

We now turn to the case with $k \neq 0$. In this case also,
the reduction that leads to an equation only for the scalar
field can be performed. However, due to the more
complex nature of the equations, this reduction no longer
leads to a second order ODE but rather to a third order one.
In particular, in this case the reduction (after
differentiating (\ref{ceq4}) and substituting the
result, as well as (\ref{ceq2}) and (\ref{ceq4}), 
in (\ref{ceq3})) results in
\begin{eqnarray}
6\dddot{\phi} \dot{\phi}-8 \ddot{\phi}^2 -10 m^2 \phi \ddot{\phi}
- 6 \dot{\phi}^4 + 6 m^2 \dot{\phi}^2 + 3 m^2 \phi^2 \dot{\phi}^2
- 2 m^4 \phi^2 =0 
\label{ceq10}
\end{eqnarray}
It is interesting to note then that the massless case once again
shares the exact same, finite time blowup solutions of Eq.
(\ref{ceq8}). One can then once again use the same argument for
the massive case to identify such solutions as the dominant
asymptotic behavior for $m \neq 0$, since these terms blowup as
$(t^{\star}-t)^{-4}$ in Eq. (\ref{ceq10}), while the rest of the
terms diverge with a rate  of (at most) $O(t^{\star}-t)^{-2}$.
Numerical integration of Eq. (\ref{ceq10}) for various initial
conditions and various masses confirms the theoretical prediction
of finite time collapse. It appears however that the mass of the
scalar affects the time at which collapse will occur ($t^{\star}$)
(see e.g., Fig. \ref{fmfig2}).

It is also worth noting that in the context of Eq. (\ref{ceq10}),
the steady state $\phi=0$ appears to be linearly stable since for small
perturbations  $\sim \epsilon \exp(\omega t)$, collecting the
leading order behavior (O($\epsilon^2$)), one obtains the
algebraic stability equation of the form:
\begin{eqnarray}
(\omega^2+m^2)^2=0.
\label{ceq11}
\end{eqnarray}
Eq. (\ref{ceq11}) has double imaginary roots:
\begin{eqnarray}
\omega=\pm i m
\label{ceq12}
\end{eqnarray}
which, in turn, denote marginal stability. Notice that the
same situation occurs in the critical case of Eq. (\ref{ceq9}).
However, the nonlinear term of the form $-6 \dot{\phi}^4$ 
asymptotically dominates and gives rise to a nonlinearity-induced
instability leading to  logarithmic blowup.
This (blowup),
in fact, is to be expected since the incompleteness of geodesics
in $3+1$-dimensional spacetimes \cite{hawking}, established using
general techniques of differential topology in
\cite{hawking2,tipler}, has been shown under quite general
conditions \cite{clarke} to lead to divergence of physically
observable properties.

In conclusion, in this work, we have examined the coupling of
$3+1$-dimensional gravity to a scalar field satisfying a linear
Klein-Gordon equation. We have found that it is possible to {\it
decouple} the equation for the massive scalar from the one for the
scale factor of the FRW metric used in this work at the
``expense'' of obtaining a nonlinear ordinary differential
equation for the scalar. The ``memory'' of the coupling to gravity
has been encapsulated in the nonlinear nature of the resulting
equation. Closed form solutions of the resulting equation can be
obtained in the massless case and exhibit logarithmic divergence
of the scalar field as a function of time (and power-law vanishing
of the scale factor). It is then observed that these solutions
persist as dominant asymptotic behavior in the case where $m \neq
0$. These results are corroborated by numerical integration of the
nonlinear ODEs in the massive case. Even though slightly different
methods have been used for the cases where the metric is flat
($k=0$) and when $k \neq 0$, the same principal conclusions have
been drawn in all  cases.

This phenomenon of decoupling persists even in the case of more
general, anisotropic Bianchi models \cite{theod1}. 
In all these cases, special
solutions to the ensuing decoupled non-linear ODE exist and they
also exhibit blowup behavior.

Finally, we return to the analogy of the logarithmic focusing of
the solutions in this simple model (where spatial dependence was a
priori fixed in the metric) with the logarithmic blowup in the
critical case of a prototypical nonlinear partial differential
equation, namely the nonlinear Schr{\"o}dinger equation, that
sustains focusing solutions. It would be particularly interesting
to examine whether the inclusion of spatial dependence can lead to
an equation of this type (through appropriate envelope wave
expansions, given that NLS is the envelope equation for nonlinear
wave equations of the Klein-Gordon type). In particular, if, in
the presence of the spatial dependence, the reduction to a {\it
nonlinear partial differential equation} for the scalar field
provides a nonlinear KG equation, then it will be natural to
expect that the reduction to NLS and the ensuing focusing
solutions will carry through. 
If such a program succeeds, it may 
appear natural to consider the NLS as prototypical model for
a ``nonlinear quantum cosmology''.

\vspace*{0.5cm} 
G.O. Papadopoulos is currently a scholar of the Greek State
Scholarships Foundation (I.K.Y.) and acknowledges the relevant
financial support.
T Christodoulakis and G.O. Papadopoulos, acknowledge support by
the University of Athens, Special Account for the Research
Grant-No. 70/4/5000
This work was also partially supported by a University of Massachusetts
Faculty Research Grant and NSF-DMS-0204585 (PGK), as well
as the AFOSR (Dynamics and Control, IGK).


\end{document}